Research and Applications

# Geo-clustered chronic affinity: pathways from socio-economic disadvantages to health disparities

Eun Kyong Shin[1,2], Youngsang Kwon[3], and Arash Shaban-Nejad[1]

[1]Department of Pediatrics, The University of Tennessee Health Science Center – Oak-Ridge National Laboratory (UTHSC-ORNL), Center for Biomedical Informatics, Memphis, Tennessee, USA, [2]Department of Sociology, Korea University, Seoul, South Korea and [3]Department of Earth Science, The University of Memphis, Memphis, Tennessee, USA

Corresponding Author: Arash Shaban-Nejad, PhD, MPH, Le Bonheur Children's Hospital, Center for Biomedical Informatics, Department of Pediatrics, University of Tennessee Health Science Center, 50 N Dunlap St., R492, Memphis, TN 38103, USA; ashabann@uthsc.edu

## ABSTRACT

**Objective:** Our objective was to develop and test a new concept (affinity) analogous to multimorbidity of chronic conditions for individuals at census tract level in Memphis, TN. The use of affinity will improve the surveillance of multiple chronic conditions and facilitate the design of effective interventions.
**Methods:** We used publicly available chronic condition data (Center for Disease Control and Prevention 500 Cities project), socio-demographic data (US Census Bureau), and demographics data (Environmental Systems Research Institute). We examined the geographic pattern of the affinity of chronic conditions using global Moran's I and Getis-Ord Gi* statistics and its association with socio-economic disadvantage (poverty, unemployment, and crime) using robust regression models. We also used the most common behavioral factor, smoking, and other demographic factors (percent of the male population, percent of the population 67 years, and over and total population size) as control variables in the model.
**Results:** A geo-distinctive pattern of clustered chronic affinity associated with socio-economic deprivation was observed. Statistical results confirmed that neighborhoods with higher rates of crime, poverty, and unemployment were associated with an increased likelihood of having a higher affinity among major chronic conditions. With the inclusion of smoking in the model, however, only the crime prevalence was statistically significantly associated with the chronic affinity.
**Conclusion:** Chronic affinity disadvantages were disproportionately accumulated in socially disadvantaged areas. We showed links between commonly co-observed chronic diseases at the population level and systematically explored the complexity of affinity and socio-economic disparities. Our affinity score, based on publicly available datasets, served as a surrogate for multimorbidity at the population level, which may assist policy-makers and public health planners to identify urgent hot spots for chronic disease and allocate clinical, medical and healthcare resources efficiently.

## INTRODUCTION

Chronic diseases impose heavy burdens on health systems, economies, and societies.[1] About half of all Americans live with at least one chronic condition and more than 75% of healthcare cost is associated with people with chronic diseases.[2] Furthermore, 31.5% of the population in the United States currently experience multiple long-term disorders.[3] Having one chronic condition is qualitatively different in nature from living with two or more chronic conditions simultaneously.[4] To complicate matters, chronic diseases often have complex causal pathways and unanticipated consequences.[5] Dealing with multimorbidity, the presence of two or more chronic conditions, often requires complex and holistic care and the use of multiple medications (polypharmacy).[6] Multimorbidities are also associated with a wide range of disadvantages; for example, elderly populations with multiple chronic conditions have a shorter life expectancy than their peers with a single disease.[7]

A growing body of research exists on the patterns of multimorbidity[6,8–10] and the social determinant of health outcomes.[11,12] Recently, the concept of multimorbidity has been expanded to encompass social-environmental factors as opposed to the isolated, single physiological-unit per individuals approach.[8,13] The manifestation of multimorbidity arises within the social context of an individual.[8] The distribution of chronic disease prevalence is closely related to socio-economic deprivation.[9] Moreover, the distribution of multimorbidity of chronic conditions inevitably varies over different geographic areas and neighborhoods.[14–17]

From the public health surveillance point of view, the main challenge is in securing accurate multimorbidity data at the population level, which would require access to all medical records of individuals in the target area. Hence, to retain patient privacy and lessen data acquisition requirements, in this study, we introduce the concept of affinity in chronic diseases as a surrogate for multimorbidity at the population level. As analogous to multimorbidity for individuals, affinity in chronic conditions measures the degree to which many chronic conditions are simultaneously prevalent in a geographical unit and thus indicate urgent surveillance hot spots for priority in clinical, medical, and healthcare resource allocations.

We chose Memphis, Tennessee for our empirical case study. Memphis is one of the poorest major metropolitan areas in the United States.[18] and its population suffers greatly from numerous health problems including a high prevalence of obesity, asthma, and diabetes. According to the 2015 health assessment report, most of the chronic disease prevalence in Memphis exceeds the national average.[19] Different stakeholders, including public health organizations in Memphis, require a better understanding of the multimorbidities and their underlying causes to design effective health interventions.

Despite the importance of multimorbidity distribution patterns of chronic diseases, those patterns have not been rigorously investigated at the population level. Little is known about how major chronic conditions are inter-related or their distribution across diverse neighborhoods. Thus, the purpose of this study was 2-fold. First, we investigated if there exists any disproportionate distribution of chronic affinity prevalence and any relationships among major chronic conditions. Second, by integrating chronic disease data and socio-economic dataset in Memphis, we scrutinized whether chronic affinity prevalences had a distinctive geographic pattern and if so, to what extent that disproportionate multimorbidity related to socio-economic disadvantages such as crime,[14,20] poverty,[6,21] and unemployment,[22] which are known to be critical socio-economic factors in explaining the chronic condition prevalence. To further validate the link between social disadvantage and coprevalences in chronic conditions, we examined whether these social factors are significant even after controlling for smoking prevalence, a known major negative behavioral contributor for most chronic conditions.[23]

In this study, we focus on the six major chronic conditions, as suggested by the US Center for Disease Control and Prevention (CDC): namely heart diseases, stroke, diabetes, arthritis, obesity, and respiratory diseases in Memphis.

## DATA AND METHODS

We used well-known, publicly available data to gauge comorbidity patterns in Memphis, Tennessee. Chronic conditions and smoking prevalence data were taken from the CDC 500 cities dataset (2015). Socio-economic indicators, poverty, unemployment, and demographic information were downloaded from the American Community Survey (2011–2015), and the total crime index was obtained from the Environmental Systems Research Institute Demographics data (2010–2017).[24] Data from multiple sources have been merged at the census tract level. Descriptive statistics of the variables are summarized in Table 1.

We calculated the correlation matrix to examine inter-relations among the six major chronic conditions. We then used the Pearson correlation coefficient to see if there was any linear correlation between any two chronic conditions. Then to unpack the multiple chronic condition prevalence pattern in Memphis, we first devised an affinity score by counting the total number of chronic conditions that exhibit a higher prevalence than the mean prevalence in Memphis. The affinity score (hereafter affinity) indicates the level of exceeding multiple chronic condition prevalence. Because we expected the chronic conditions under study may exhibit different relations to other chronic conditions and diseases, we calculated affinity for each chronic condition. In practice, as we do not have representative health data at the individual level, alternatively, we created a dummy variable for each chronic condition per census tract level, indicating exceeding prevalence of a chronic condition compared to other areas in Memphis. For example, at the census tract level, where asthma and obesity prevalence are higher than the mean value of asthma and obesity in Memphis, with all other chronic diseases under the mean value, the high prevalence of multiple chronic conditions in the area has two conditions (ie, asthma and obesity) with a chronic affinity of two. Affinity can vary from zero to six. Low affinity means that there are weak relations among the major chronic conditions and high affinity indicates that there are strong relations between them.

Two sets of empirical tests were conducted. First, we tested if there was any distinctive geographic pattern in affinity distribution in Memphis. We examined the geographic association of the affinity by measuring the spatial autocorrelation of affinity (ie, detection of clustering) using global Moran's I statistic and further investigated whether high or low values of affinity are locally clustered over the study area with Getis-Ord Gi* statistic.[25] Next, using robust regression analyses, we tested if there is any statistically significant association between socio-economic disadvantages—crime, poverty and unemployment—and affinity prevalence at the census tract level ($N = 178$). We used robust multivariate regression analyses to minimize the distortive effect of outliers as some chronic conditions have significant outliers and to account for possible heteroscedasticity (the circumstance in which the variability of a variable is unequal across the range of values of a second variable that predicts it).[26] We tested the associations between socio-economic disadvantages and chronic affinity. Using affinity as a dependent variable, in the first model, we looked at exclusively socio-economic disadvantages. As for independent variables for social conditions, we used poverty, unemployment, and crime in the neighborhood to capture the social environments within the census tract level. In the second model, we added a high-risk behavioral factor, smoking, to test if social conditions remained significant after controlling for a behavioral pattern. Given the fact that chronic condition prevalences can be sensitive for age and demographic distribution, both models included these demographic factors as control variables: Percent of the male population, percent of the population 67 years, and over and total population size.

**Table 1.** Descriptive statistics of variables at the US census tract level

| Variable names | Operationalization | Mean | Std. Dev. |
| --- | --- | --- | --- |
| Arthritis | Model-based estimate for crude prevalence of arthritis among adults aged $\geq 18$ years | 26.776 | 5.780 |
| Asthma | Model-based estimate for crude prevalence of current asthma among adults aged $\geq 18$ years | 10.961 | 1.776 |
| Diabetes | Model-based estimate for crude prevalence of diagnosed diabetes among adults aged $\geq 18$ years | 15.642 | 5.962 |
| Heart disease | Model-based estimate for crude prevalence of coronary heart disease among adults aged $\geq 18$ years | 7.388 | 2.558 |
| Obesity | Model-based estimate for crude prevalence of obesity among adults aged $\geq 18$ years | 38.544 | 7.991 |
| Stroke | Model-based estimate for crude prevalence of stroke among adults aged $\geq 18$ years | 4.878 | 2.254 |
| Affinity | Total number of chronic conditions that exceed the mean value of Memphis | 2.960 | 2.584 |
| Crime | Environmental Systems Research Institute Total Crime Index | 47.970 | 33.420 |
| Poverty | Percentage of people living below the federal poverty line living in the census tract | 28.864 | 16.624 |
| Unemployment | Percent of unemployed population living in the census tract | 15.729 | 9.315 |
| Smoking | Current smoking among adults aged $\geq 18$ years | 25.376 | 7.194 |
| Percentage of population 67 years and older | Percent population 67 and over in the census tract | 9.135 | 4.959 |
| Percentage of male population | Percent male population in the census tract | 48.073 | 5.920 |
| Population | Total number of people living in the census tract (US Census 2010) | 3634.107 | 1718.732 |

## RESULTS

The major chronic conditions in Memphis were closely linked (Table 2). The Pearson's correlation among the six major chronic conditions were all above 0.5 ($P < 0.001$) with the highest correlation ($R = 0.98$) found between diabetes and stroke followed by arthritis and heart disease ($R = 0.96$). The lowest correlation ($R = 0.54$) was found between asthma and arthritis. About 35% of census tracts in Memphis exhibited all six chronic conditions. The multimorbidity affinity map (Figure 1A) shows distinctive disproportionate patterns of multimorbidity. High affinity ($> 5$) were concentrated in the west side of Memphis except for a few census tracks with zero affinity. Global Moran's I (Z score of 17.8, $P < 0.001$) confirmed a statistically significant overall cluster of multimorbidity over the study area. Getis-Ord Gi* statistics further identified local hot spots (ie, clusters of high values of affinity) near the downtown area (where poverty is concentrated) while cold spots (ie, clusters of low values of affinity) were located mainly east, outward of Memphis (Figure 2). The correlation between poverty (Figure 1B) and affinity (Figure 1A) was 0.679 and between unemployment (Figure 1C) and affinity was 0.675. The correlation between crime (Figure 1D) and affinity was 0.508. In the next regression analyses, we investigated why we observed this west-bound, skewed distribution to further examine if the underlying socio-economic disparities were significantly associated with this disproportionate affinity prevalence.

Our robust regression results (Table 3) showed statistically significant links between the socio-economic conditions and affinity. Per multicollinearity, we checked the variance inflation factor (VIF). The mean value of VIF was 1.73 and we didn't have any variable that exceeds 3 or falls below 1.00. Also, our model passed the linktest for specification error. As shown in Model 1, Table 3, poverty was significantly associated with increased affinity and unemployment was associated with an 0.081 affinity increase. The crime prevalence was significantly associated with the chronic affinity. If a neighborhood suffered more from poverty, high unemployment rate and crime it was also more likely to exhibit higher affinity between multiple chronic conditions even after controlling within the census tract level for the older population proportion, male population, and total population. Social disparities of poverty, unemployment, and crime indicate not only the social insecurity of a neighborhood but also a higher association among chronic conditions of arthritis, asthma, diabetes, heart disease, obesity, and stroke. The disadvantages are not only social but also medical. In addition, when we added a behavioral component of smoking, to Model 1, the results further confirmed the association between affinity and crime (Model 2, Table 3). With the inclusion of smoking, the health disadvantages continued to disproportionately cling to socially disadvantaged areas with higher crime rates, even after controlling for smoking prevalence and other demographic factors. Confirming the existing literature, smoking prevalence was positively associated with the affinity. The crime prevalence was positively associated with the chronic affinity. As for poverty and unemployment, effect directionalities remained positive but they lost statistical significance after controlling for smoking prevalence.

These results could be driven by a correlation between smoking and poverty and/or unemployment in Memphis, TN.

## DISCUSSION

Given the high prevalence and related healthcare costs, preventing chronic conditions is one of the major priorities for public health organizations. Multimorbidity, with its additional layers of complexity and complication, has increased and the trend is expected to continue across the United States.[27] Knowing how different chronic conditions are related to other chronic conditions helps public health practitioners and policymakers to design, implement, and deliver effective health programs and interventions. In this article, we have shown multiple chronic condition prevalence is related to socio-economic conditions. An affinity score in chronic diseases, which is constructed from publicly available datasets, serves as a practical surrogate for multimorbidity at the population level. Such an affinity score can assist policymakers and public health organizations to efficiently identify urgent hot spots. For instance, if the Memphis city council plans to build a health center, having it located in the west part of Memphis can improve the accessibility of the most vulnerable population to healthcare and therefore improve the overall population's chronic health conditions more efficiently. In addition, depending on the intervention area, health educators and policymakers can design geo-tailored health programs and link the health intervention program with existing social work interventions as applicable. For instance, in the case of designing a health intervention in a neighborhood suffering significantly from both high asthma and high obesity, communications and collaborations between experts in related areas should be encouraged and rigorously incorporated.

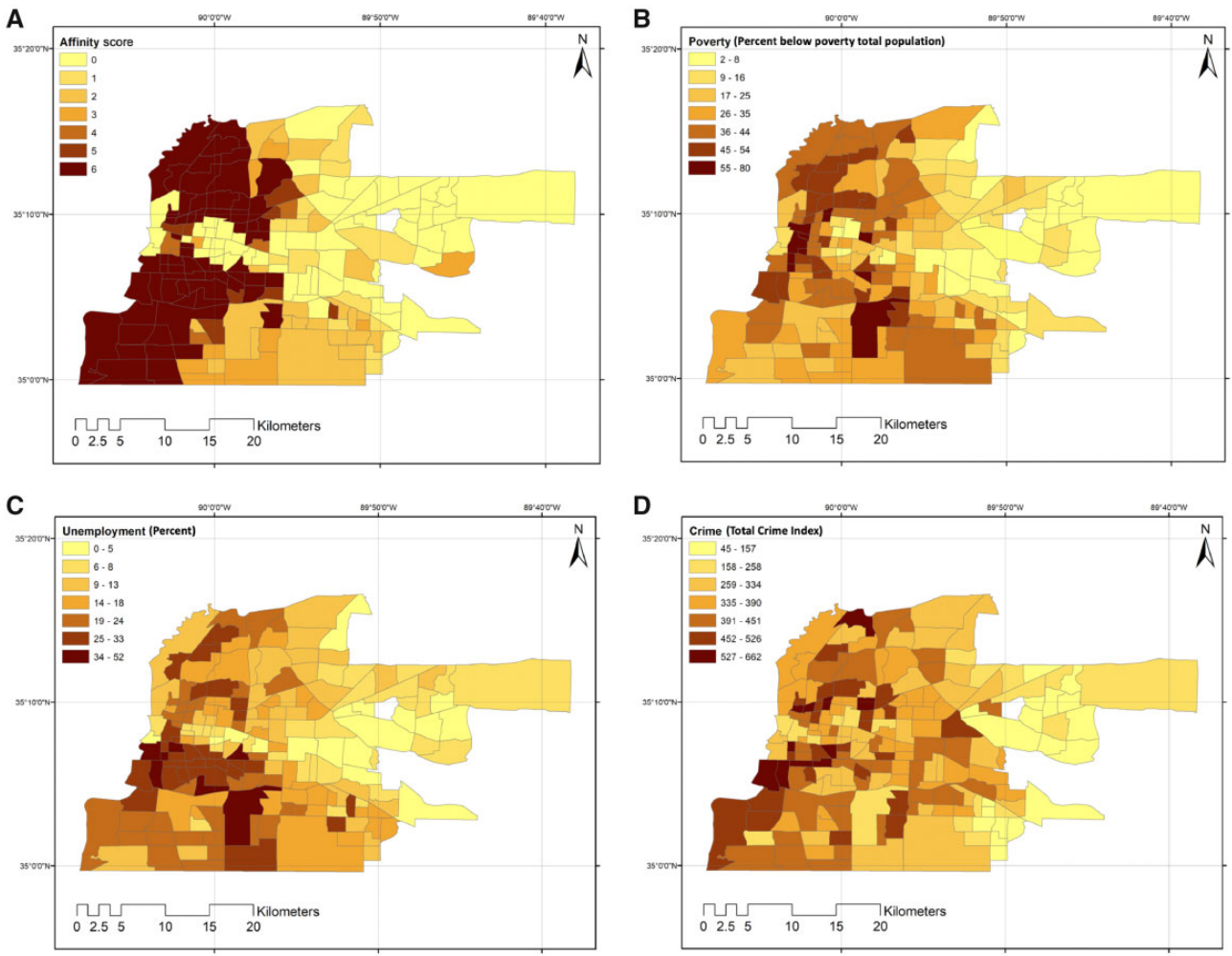

Figure 1. Distribution of socio-economic variables and chronic affinity in Memphis. A. Chronic affinity. B. Poverty. C. Unemployment. D. Crime: The crime index values are all referenced by 100, representing US average crime.

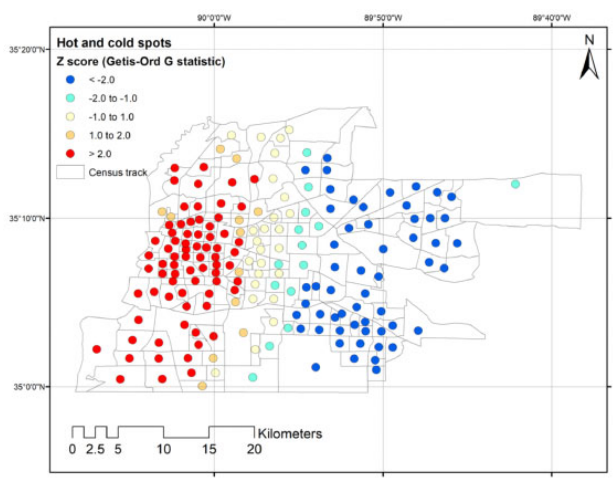

Figure 2. Hot spot analysis using Getis-Ord G* Statistic.

Table 2. Pearson's correlations among major chronic condition prevalences (sig at 0.001)

|  | Arthritis | Asthma | Heart disease | Diabetes | Obesity | Stroke |
|---|---|---|---|---|---|---|
| Arthritis | – | | | | | |
| Asthma | 0.54 | – | | | | |
| Heart disease | 0.96 | 0.59 | – | | | |
| Diabetes | 0.86 | 0.82 | 0.90 | – | | |
| Obesity | 0.58 | 0.94 | 0.64 | 0.88 | – | |
| Stroke | 0.89 | 0.78 | 0.95 | 0.98 | 0.83 | – |

In addition, we found that the accumulation of chronic conditions, measured by chronic affinity, was unevenly distributed throughout Memphis. Not only socio-economic disadvantages were significantly linked to higher coprevalence in major chronic conditions, but coprevalence of major chronic conditions were also heavily clustered in socially disadvantaged neighborhoods. The socially deprived area suffers more from multiple chronic conditions. More crime, severe poverty and high unemployment lead to a higher likelihood of close links among arthritis, asthma, diabetes, heart disease, obesity, and stroke. Social conditions can influence the health culture within the neighborhood. For example, a high crime prevalence may lead to limited physical activities in the neighborhood and thus affect the culture of health negatively.

Table 3. Robust regression results for affinity scores

|  |  | Model 1 | | | Model 2 | | |
|---|---|---|---|---|---|---|---|
|  |  | Coef. | P | [95% CI] | Coef. | P | [95% CI] |
| Independent variables | Poverty | 0.067 | 0.000 | 0.037, 0.096 | 0.000 | 0.995 | −0.037, 0.037 |
|  | Unemployment | 0.081 | 0.007 | 0.023, 0.138 | 0.031 | 0.230 | −0.020, 0.082 |
|  | Crime | 0.003 | 0.024 | 0.000, 0.006 | 0.002 | 0.034 | 0.000, 0.005 |
|  | Smoking | – | – | – | 0.240 | 0.000 | 0.144, 0.337 |
| Control variables | Male population | −0.039 | 0.204 | −0.100, 0.021 | −0.053 | 0.028 | −0.101, −0.006 |
|  | 67 years and over population | 0.098 | 0.001 | 0.044, 0.153 | 0.164 | 0.000 | 0.110, 0.218 |
|  | Total population | 0.000 | 0.389 | 0.000, 0.000 | 0.000 | 0.772 | 0.000, 0.000 |

The present study has some limitations and empirical constraints. Our study looks at one time-point in one urban area. To better understand the dynamics of clusters among multiple chronic conditions, investigations of longitudinal trends in multiple places are needed. Additionally, our health indicators do not provide detailed age break down. Among the top major chronic conditions, cancer is not included in this study, due to its high sensitivity to genetic factors[28] and its higher prevalence in the elderly population.[29] Despite these constraints, this study shows a geo-spatial link and dependency between social disadvantages and chronic affinity. Medical and health disadvantages were disproportionately accumulated in socially disadvantaged neighborhoods. To minimize the potential adverse effects of coprevalence of multiple chronic conditions, the geo-distinctive pattern of chronic affinity should be considered and implemented in the health surveillance system. The affinity score can facilitate efficient chronic disease controls and timely response as well as public health planning and decision making. Particularly, when policymakers and public health service providers have limited access to individual medical records in a targeted area, affinity scores can provide actionable evidence.

## FUNDING STATEMENT

The research has been funded by the University of Tennessee Health Science Center.

## CONTRIBUTORSHIP STATEMENT

E.K.S. designed the study, curated the datasets, conducted the statistical analysis and drafted the manuscript. E.K.S. had full access to all the data in the study and takes responsibility for the integrity of the data and the accuracy of the data analysis. Y.K. contributed to the study design and data curation. Y.K. conducted the geo-analysis, mapping and contributed in writing the manuscript for the methods and results. A.S.N conceived the study, participated in the study design, writing and revising the manuscript, obtained funding, and provided administrative support. All authors read and approved the final manuscript.

## COMPETING INTERESTS STATEMENT

Authors do not have competing interests.